\documentclass[reqno]{amsart}
\usepackage{hyperref}
\usepackage{xcolor}
\usepackage{amsthm,amsmath,amssymb}
\usepackage{mathrsfs}
\usepackage{cleveref}
\usepackage{wasysym}
\usepackage[T1]{fontenc}
\usepackage{inputenc}[utf8]
\usepackage{enumerate}

\renewcommand{\d}{\mathrm{d}}

\renewcommand{\epsilon}{\varepsilon}

\renewcommand{\eth}{\text{\rm{\dh}}}
\renewcommand{\thorn}{\text{\rm{\th}}}

\newcommand{\scri}{\mathscr{I}}

\DeclareFontFamily{U}{BOONDOX-calo}{\skewchar\font=45 }
\DeclareFontShape{U}{BOONDOX-calo}{m}{n}{
  <-> s*[1.05] BOONDOX-r-calo}{}
\DeclareFontShape{U}{BOONDOX-calo}{b}{n}{
  <-> s*[1.05] BOONDOX-b-calo}{}
\DeclareMathAlphabet{\mathcalboondox}{U}{BOONDOX-calo}{m}{n}
\SetMathAlphabet{\mathcalboondox}{bold}{U}{BOONDOX-calo}{b}{n}
\DeclareMathAlphabet{\mathbcalboondox}{U}{BOONDOX-calo}{b}{n}

\begin{document}
\title[Scattering of Maxwell Potentials]
{Scattering of Maxwell Potentials on Curved Spacetimes}

\author[Grigalius Taujanskas\hfil \hfilneg] {Grigalius Taujanskas}  

\address{Grigalius Taujanskas \newline
Trinity Hall \newline Trinity Lane \newline Cambridge CB2 1TJ}
\email{taujanskas@dpmms.cam.ac.uk}

\subjclass[2000]{} 
\begin{abstract}
We report on the recent construction of a scattering theory for Maxwell potentials on curved spacetimes \cite{TaujanskasNicolas2022}. \end{abstract}

\maketitle \numberwithin{equation}{section}
\newtheorem{theorem}{Theorem}
\newtheorem*{theorem*}{Theorem}
\newtheorem{corollary}[theorem]{Corollary}
\newtheorem{lemma}[theorem]{Lemma}
\newtheorem{remark}[theorem]{Remark}
\newtheorem{problem}[theorem]{Problem}
\newtheorem{example}[theorem]{Example}
\newtheorem{definition}[theorem]{Definition}
\allowdisplaybreaks

\section{Introduction}

The study of the asymptotic structure of isolated systems in general relativity has been a rich area of research since at least the 1960s. A number\footnote{Too many to list here, see for example \cite{HintzVasy2017} for a more complete bibliography.} of landmark results \cite{Friedrich86,Penrose1965,ChristodoulouKlainerman1990,HintzVasy2017} have been established, however many important questions, particularly regarding the fine asymptotic properties of fields and the rigorous analytic formulations of scattering theories, remain. In particular, it is of interest to study the far-field regime of massless fields---such as gravity or electromagnetism---on curved spacetimes, where they are scattered by background curvature. Since massless fields enjoy an essential conformal invariance, Penrose's conformal method \cite{Penrose1965} provides an excellent conceptual framework to study their scattering and asymptotics.

In \cite{TaujanskasNicolas2022} the author and J.-P. Nicolas construct a complete scattering theory for Maxwell potentials on a class of curved, non-stationary spacetimes. The scattering construction of \cite{TaujanskasNicolas2022} in principle allows for reasonably general backgrounds: they may contain matter fields\footnote{In principle our scattering construction allows for matter fields provided they decay sufficiently fast at infinity, specifically that $\hat{g}_{ab} \hat{\Box} \Omega \approx 4 \hat{\nabla}_a \hat{\nabla}_b \Omega $ and $\hat{\Psi}_{0,1,2,3,4} \approx 0$, where $\approx$ denotes equality on $\scri$. For simplicity, here we report on the vacuum case.}, as long as the conformal boundary is suitably smooth and the spacetime is sufficiently close to Minkowski space\footnote{Proximity to Minkowski space allows one to construct a concrete and fairly large class of spacetimes on which the scattering theory of \cite{TaujanskasNicolas2022} holds. However, this is likely not strictly necessary provided the spacetime has the correct Penrose diagram, and there may be examples of `large' allowable background spacetimes.}. In the case of \emph{vacuum}\footnote{In fact, the authors of \cite{ChruscielDelay2003} comment that their constructions should apply to certain non-vacuum constraint equations, e.g. the Einstein--Maxwell system.}, a concrete subclass of such spacetimes---to which we refer as Corvino--Schoen--Chru\'sciel--Delay (CSCD) spacetimes---may be constructed using the initial data gluing theorems of \cite{ChruscielDelay2003,Corvino2000,CorvinoSchoen2006,ChruscielDelay2002} for the vacuum Einstein equations, and Friedrich's theorem for the semi-global stability of Minkowski space \cite{Friedrich86,FriedrichSchmidt1987}. This produces an infinite-dimensional family of vacuum spacetimes which have good conformal compactification properties: they are asymptotically simple in the sense of Penrose \cite{Penrose1965}, their null and timelike infinities can be ensured to be $C^k$ for any integer $k$, and they are exactly Schwarzschildean, or Kerrian, in a neighbourhood of spatial infinity. For simplicity, we work with the case of exactly Schwarzschildean spacetimes near $i^0$. CSCD spacetimes are described in more detail below.

\section{Background Spacetimes and Field Equations}

\subsection{CSCD Spacetimes}
\label{sec:CSCD_spacetimes}

We work on spacetimes $(\mathcalboondox{M}, g_{ab})$ which are four dimensional, globally hyperbolic, asymptotically flat, and arise as developments of the vacuum Einstein equations
\[ \mathrm{Ric}(g) = 0 \]
from initial data $(h_{ab}, \kappa_{ab})$ on a Cauchy hypersurface $\Sigma \simeq \mathbb{R}^3$ such that:
\begin{enumerate}[(i)]
\setlength\itemsep{3pt}
	\item  outside a given compact set $K \subset \Sigma$, the data $(h_{ab}, \kappa_{ab}) = (g_{ab}^{\text{Schw}}(t=0), 0)$ is exactly Schwarzschild at $t=0$, so that the development $(\mathcalboondox{M}, g_{ab})$ is exactly Schwarzschild in a neighbourhood of spatial infinity $i^0$,
	\item the data $(h_{ab}, \kappa_{ab})$ is sufficiently close to Minkowskian data in the sense required by the theorems of, say, \cite{ChruscielDelay2003},
	\item the initial metric $h_{ab}$ satisfies the condition that $\| r^2 \mathrm{Ric}(h) \|_{L^\infty(\Sigma)} $ is not too large\footnote{Condition (iii) is not part of the assumptions of the theorems of Corvino, Schoen, Chru\'sciel and Delay, but comes from the construction of the initial potential. See \cref{sec:spaces_of_data}.}, for $r$ an appropriately defined radial coordinate on $\Sigma$ which coincides with the standard Schwarzschildean radial coordinate on $\Sigma \setminus K$.
\end{enumerate}
With these conditions, $(\mathcalboondox{M}, g_{ab})$ is then asymptotically simple with a $C^k$ conformal compactification (for some $k$ sufficiently large) at $\scri^\pm$ and $i^\pm$. Being exactly Schwarzschild in $D^+(\Sigma \setminus K)$, at $i^0$ the spacetime is conformally singular. We denote the conformally rescaled metric by $\hat{g}_{ab} = \Omega^2 g_{ab}$, where $\Omega$ is the corresponding conformal factor.

\subsection{Field Equations}

Maxwell's equations are conformally invariant and are given by
\[ \nabla^a F_{ab} = 0 = \nabla_{[a} F_{bc]} \iff \hat{\nabla}^a \hat{F}_{ab} =0 = \hat{\nabla}_{[a} \hat{F}_{bc]} , \]
where $\nabla$ is the Levi-Civita connection of $g_{ab}$ and $\hat{\nabla}$ is the Levi-Civita connection of $\hat{g}_{ab}$, and $\hat{F}_{ab} = F_{ab}$, i.e. $F_{ab}$ has conformal weight zero. In terms of the potential the equations read
\begin{equation} \label{Maxwell_potentials_no_gauge} \Box A_a - \nabla_b (\nabla_a A^a) + \mathrm{R}_{ab} A^a = 0, \end{equation}
and, without a choice of gauge, are also conformally invariant provided $A_a$ is chosen to have conformal weight zero. We choose an NP tetrad $(l^a, m^a, \bar{m}^a, n^a)$ on $\mathcalboondox{M}$ and the conformal scaling $(\hat{l}^a, \hat{m}^a, \bar{\hat{m}}^a, \hat{n}^a) = (\Omega^{-2} l^a, \Omega^{-1} m^a, \Omega^{-1} \bar{m}^a, n^a)$ so that on $\scri^+$ the vector field $n^a$ becomes a generator of $\scri^+$, and define the components of $A_a$ and $F_{ab}$
\[ \left( \begin{array}{ccc} A_0 & A_1 & A_2 \\ F_0 & F_1 & F_2 \end{array} \right) = \left( \begin{array}{ccc} A_a l^a & A_a n^a & A_a m^a \\ F_{ab} l^a m^b & \frac{1}{2} F_{ab} (l^a n^b + \bar{m}^a m^b ) & F_{ab} \bar{m}^a n^b \end{array} \right), \]
with the associated conformal weights inherited from the scaling of the tetrad. Moreover, we choose a uniformly timelike vector field $T^a$, which in the case of Minkowski space is exactly $T^a = \partial_t$ and in the general case coincides with the Schwarzschildean Killing vector field $\partial_t$ in a neighbourhood of $i^0$; we denote by $\mathbf{A}$ the projection of $A_a$ to hypersurfaces orthogonal to $T^a$, and write $\mathfrak{a} = T^a A_a$.

\section{Main Results}

The main results of \cite{TaujanskasNicolas2022} can be summarised in the following theorems. Our tetrad is adapted to future null infinity, so the following results are explicitly stated only in the case of $\scri^+$. The analogous gauge conditions and function spaces on $\scri^-$ can be obtained by interchanging the vector fields $l^a$ and $n^a$.

\begin{theorem} \label{thm:main_thm_Minkowski} Let $(\mathcalboondox{M}, g_{ab}) = (\mathbb{R}^4, \eta_{ab})$ be the Minkowski spacetime. Then a finite energy solution to \eqref{Maxwell_potentials_no_gauge} admits the gauge 
\begin{equation} \label{flat_gauge} \nabla_a A^a = \boldsymbol{\nabla} \cdot \mathbf{A} = \mathfrak{a} = 0,	
\end{equation}
 and there exist bounded, invertible linear operators
\begin{align*} \mathfrak{T}^\pm_K : \dot{H}^1_C(\Sigma) \oplus L^2_C (\Sigma) &\longrightarrow \dot{\mathcal{H}}^1(\scri^\pm) \\
(\mathbf{A}, \dot{\mathbf{A}})|_{\Sigma} &\longmapsto (\hat{A}_0^\pm, \hat{A}_1^\pm, \hat{A}_2^\pm),
\end{align*}
corresponding to the future/past development according to \eqref{Maxwell_potentials_no_gauge} in the gauge \eqref{flat_gauge} on $\mathcalboondox{M}$, which map finite-energy Maxwell potential initial data on $\Sigma$ to finite-energy Maxwell potential characteristic data on $\scri^\pm$. The function spaces above are given by
\begin{align*}
	\dot{H}^1_C(\Sigma) &= \{ \mathbf{A} \in \dot{H}^1(\Sigma; \mathbb{R}^3) \, : \, \boldsymbol{\nabla} \cdot \mathbf{A} = 0 \}, \\
	L^2_C(\Sigma) &= \{ \dot{\mathbf{A}} \in L^2(\Sigma; \mathbb{R}^3) \, : \, \boldsymbol{\nabla} \cdot \dot{\mathbf{A}} = 0 \},
\end{align*}
and
\begin{align*} \dot{\mathcal{H}}^1(\mathscr{I}^+) &= \bigg\{ (\hat{A}^+_0, \hat{A}_1^+, \hat{A}^+_2) \, : \, \hat{A}_0^+ = \int_{-\infty}^u 2 \operatorname{Re} \hat{\eth} \bar{\hat{A}}_2^+ \, \d u, ~ \hat{A}_1^+ = 0, \\
	& \qquad \int_{\scri^+} | \partial_u \hat{A}_2^+ |^2 \, \d u \wedge \mathrm{dv}_{\mathbb{S}^2} < \infty \bigg\} \\
	& \simeq \dot{H}^1(\mathbb{R}; L^2(\mathbb{S}^2)),  	
\end{align*}
and analogously for $\dot{\mathcal{H}}(\scri^-)$. Consequently, there exists a bounded, invertible linear scattering operator
\begin{align*} \mathscr{S}_K = \mathfrak{T}^+_K \circ (\mathfrak{T}_K^-)^{-1} : \dot{\mathcal{H}}^1(\scri^-) & \longrightarrow \dot{\mathcal{H}}^1(\scri^+), \\
(\hat{A}_0^-, \hat{A}_1^-, \hat{A}_2^-) &\longmapsto (\hat{A}_0^+, \hat{A}_1^+, \hat{A}_2^+)
\end{align*}
which corresponds to the development according to \eqref{Maxwell_potentials_no_gauge} in the gauge \eqref{flat_gauge} of $(\hat{A}_0^-, \hat{A}_1^-, \hat{A}_2^-)$ from $\scri^-$. The subscript $K$ in the above refers to the standard timelike Killing field $K = \partial_t$.

Moreover, the Morawetz vector field
\[ K_0 = (t^2 + r^2) \partial_t + 2tr \partial_r, \]
gives rise to a stronger scattering theory given by bounded, invertible linear operators $\mathfrak{T}^\pm_{K_0}$, where
\[ \mathfrak{T}^+_{K_0} : r^{-1} \dot{H}^1_C(\Sigma)^{\mathrm{curl}} \oplus r^{-1} L^2_C(\Sigma) \longrightarrow u^{-1} \dot{\mathcal{H}}^1(\scri^+) \]
and similarly for $\mathfrak{T}_{K_0}^-$, where 
\begin{align*}
	r^{-1} \dot{H}^1_C(\Sigma)^{\mathrm{curl}} &= \{ \mathbf{A} \in \dot{H}^1(\Sigma; \mathbb{R}^3) \, : \, \boldsymbol{\nabla} \cdot \mathbf{A} = 0,  ~r (\boldsymbol{\nabla} \times \mathbf{A}) \in L^2(\Sigma; \mathbb{R}^3) \}, \\
	r^{-1} L^2_C(\Sigma) &= \{ \dot{\mathbf{A}} \in L^2(\Sigma; \mathbb{R}^3) \, : \, \boldsymbol{\nabla} \cdot \dot{\mathbf{A}} = 0, ~ r \dot{\mathbf{A}} \in L^2(\Sigma; \mathbb{R}^3) \},
\end{align*}
and 
\begin{align*} u^{-1} \dot{\mathcal{H}}^1(\scri^+) = \bigg\{ &(\hat{A}_0^+, \hat{A}_1^+, \hat{A}_2^+) \, : \, \hat{A}^+_0 = \int_{-\infty}^u 2 \operatorname{Re} \hat{\eth} \bar{\hat{A}}_2^+ \, \d u, ~ \hat{A}_1^+ = 0, ~ \\ &\int_{\scri^+} \left( u^2 |\partial_u \hat{A}_2^+|^2 + | \hat{\eth} \bar{\hat{A}}_2^+ |^2 \right) \d u \wedge \mathrm{dv}_{\mathbb{S}^2}< \infty  \bigg\},
\end{align*}
and similarly for $v^{-1} \dot{\mathcal{H}}^1(\scri^-)$. The resulting scattering operator 
\[ \mathscr{S}_{K_0} = \mathfrak{T}^+_{K_0} \circ (\mathfrak{T}^-_{K_0})^{-1} : v^{-1} \dot{\mathcal{H}}^1(\scri^-) \longrightarrow u^{-1} \dot{\mathcal{H}}^1(\scri^+)  \]
is linear, bounded, invertible, and maps past asymptotic data $(\hat{A}_0^-, \hat{A}_1^-, \hat{A}_2^-)$ to future asymptotic data $(\hat{A}^+_0, \hat{A}_1^+, \hat{A}_2^+)$ through a development according to \eqref{Maxwell_potentials_no_gauge} in the gauge \eqref{flat_gauge}.
\end{theorem}

\begin{theorem}
	Let $(\mathcalboondox{M}, g_{ab})$ be a CSCD spacetime as described in \cref{sec:CSCD_spacetimes}. Then a finite energy solution to \eqref{Maxwell_potentials_no_gauge} admits a gauge which satisfies the conditions
	\begin{enumerate}[(i)]
	\setlength\itemsep{4pt}
		\item $\nabla_a A^a = 0$	 in a neighbourhood of $\Sigma$ and a neighbourhood of $\scri^+$,
		\item $\mathfrak{a}|_\Sigma = 0 = \boldsymbol{\nabla} \cdot \mathbf{A}|_\Sigma$, and
		\item $\hat{A}_1^{[1]}|_{\scri^+} = 0$,
 	\end{enumerate}	
	where $\hat{A}_1^{[1]} = \Omega^{-1} \hat{A}_1$, and there exist bounded, invertible linear operators
	\begin{align*} \mathfrak{T}^\pm : \dot{H}^1_C(\Sigma)^{\mathrm{curl}} \oplus L^2(\Sigma) &\longrightarrow \dot{\mathcal{H}}^1(\mathscr{I}^\pm), \\
	(\mathbf{A}, \nabla_T\mathbf{A})|_\Sigma & \longmapsto (\hat{A}_0^\pm, \hat{A}_1^\pm, \hat{A}_2^\pm),
	\end{align*}
	corresponding to the future/past development according to \eqref{Maxwell_potentials_no_gauge} on $\mathcalboondox{M}$ in the above gauge, which map finite-energy Maxwell potential initial data on $\Sigma$ to finite-energy Maxwell potential characteristic data on $\mathscr{I}^\pm$. The function spaces above are given by
	\[ \dot{H}^1_C(\Sigma)^{\mathrm{curl}} = \{ \mathbf{A} \in \dot{H}^1(\Sigma) \, : \, \boldsymbol{\nabla} \cdot \mathbf{A} = 0, ~ \boldsymbol{\nabla} \times \mathbf{A} \in L^2(\Sigma) \} \]
	and $\dot{\mathcal{H}}^1(\scri^\pm)$ as in \Cref{thm:main_thm_Minkowski}. Consequently, there exists a bounded, invertible linear scattering operator
	\begin{align*} \mathscr{S} = \mathfrak{T}^+ \circ (\mathfrak{T}^-)^{-1}: \dot{\mathcal{H}}^1(\scri^-) &\longrightarrow \dot{\mathcal{H}}^1(\scri^+)	\\
	(\hat{A}_0^- , \hat{A}_1^-, \hat{A}_2^-) & \longmapsto (\hat{A}_0^+ , \hat{A}_1^+, \hat{A}_2^+)
	\end{align*}
	which corresponds to the development of $(\hat{A}_0^-, \hat{A}_1^-, \hat{A}_2^-)$ from $\scri^-$ according to \eqref{Maxwell_potentials_no_gauge} on $\mathcalboondox{M}$ in the above gauge.
\end{theorem}

\section{Remarks}

\subsection{Conformal scale} The construction of the gauge and the spaces of characteristic data rely on the existence of a conformal scale which satisfies a number of conditions. In effect, we construct a conformal scale in which $\scri^+$ is almost as `flat' as in the case of Minkowski space, which in general is permitted by the smoothness of the conformal boundary and the rapid decay of background matter fields at infinity. In this scale we have that the spin coefficients (cf. \cite{PenroseRindler2}) $\hat{\lambda}$, $\hat{\pi}$, $\hat{\mu}$, $\hat{\tau}$ and $\hat{\gamma}$ vanish on $\scri^+$, $\hat{\nu}$ vanishes in a neighbourhood of $\scri^+$, and $\hat{\mu}$ is real in a neighbourhood of $\scri^+$. Moreover, the components $\hat{\Phi}_{21}$ and $\hat{\Phi}_{22}$ of the trace-free Ricci tensor of $\hat{g}_{ab}$ vanish on $\scri^+$, as does the full rescaled Weyl tensor. This conformal scale is the analogue of the conformal factor $\Omega = r^{-1}$ in Minkowski space.

\subsection{Spaces of data} \label{sec:spaces_of_data}

The spaces of initial and characteristic data are derived from the (conformally covariant) Maxwell stress-energy tensor and a choice of timelike conformal Killing field, together with various gauge conditions on the potential. For the space of characteristic data (on $\scri^+$), the expression for the transverse component $\hat{A}_0^+$ comes from the reduction of our gauge to $\scri^+$. Precisely, the Lorenz gauge in the physical spacetime reduces to the condition $\hat{A}_1 = 0$ on $\scri^+$ at first order in $\Omega$, and to the condition 
\[ -f \hat{A}_1^{[1]} + \hat{\thorn}' \hat{A}_0 - 2 \operatorname{Re}\hat{\eth} \bar{\hat{A}}_2 = 0 \]
on $\scri^+$ at second order in $\Omega$, for a smooth function $f$. This becomes an ODE for $\hat{A}_0$ on $\scri^+$ if we impose the additional gauge condition that $\hat{A}_1^{[1]} = 0$ on $\scri^+$, which we then solve by integrating in the Bondi parameter $u$, $\hat{\thorn}' = \partial_u$, on $\scri^+$. Note that we set $\hat{A}_0^+$ to vanish at $i^0$. If the free data $\hat{A}_2^+$ is smooth and compactly supported, for example, then the formula for $\hat{A}_0^+$, being an integral along $\mathscr{I}^+$ of a function which decays towards both $i^+$ and $i^0$, means that $\hat{A}_0^+$ can be chosen to vanish at either $i^0$ or $i^+$, but not both. We expect the difference $\hat{A}^+_0|_{i^+} - \hat{A}_0^+|_{i^0}$ to be related to the electromagnetic memory effect. This will be explored elsewhere.

The conditions on the space of initial data are reasonably self-explanatory. What is not immediately obvious, however, is that the condition on the \emph{background spacetime} 
\begin{equation} \label{Ricci_bound_Hardy} \| r^2 \operatorname{Ric}(h) \|_{L^\infty(\Sigma)} < C^{-1} \end{equation}
for some constant $C$ in fact arises from the construction of the initial data. This is due to the following reason. On a general (e.g. CSCD) spacetime, even in the gauge \eqref{flat_gauge} on $\Sigma$, the canonical energy on $\Sigma$ does not define a norm on the potential due to the presence of the Ricci curvature of $h$. One must therefore show that there is a one-to-one correspondence between finite energy fields $\mathbf{E}, \, \mathbf{B} \in L^2(\Sigma)$ and potentials in a suitable space by some other means. Essentially, this amounts to solving the elliptic system
\[ \boldsymbol{\Delta} \mathbf{A}_i + \mathbf{R}_{ij} \mathbf{A}^j = - (\boldsymbol{\nabla} \times \mathbf{B})_i \]
on $\Sigma$. However, the Ricci curvature $\mathbf{R}_{ij} = \mathrm{Ric}(h)_{ij}$ on $\Sigma$ is in general not positive-definite, and so standard elliptic theory fails here. Indeed, it is not clear what the kernel of the operator from $\dot{H}^1(\Sigma; \mathbb{R}^3)$ into $\dot{H}^{-1}(\Sigma; \mathbb{R}^3)$, as defined by the left-hand side of the above equation, is in general. If the assumption \eqref{Ricci_bound_Hardy} on $\mathrm{Ric}(h)$ is made, however, then it is possible to ensure, using Hardy's inequality on $\Sigma$, that $\| \mathbf{A} \|_{\dot{H}^1(\Sigma)} \lesssim \| \mathbf{B} \|_{L^2(\Sigma)}$, which is then sufficient to control the regularity of the initial data.

\subsection{Goursat problem}

The invertibility of the operators $\mathfrak{T}^\pm$ is equivalent to the well-posedness of the characteristic initial value problem (or \emph{Goursat} problem) for \eqref{Maxwell_potentials_no_gauge} from $\scri^\pm$ with finite energy characteristic data. The main analytic tool that enables us to solve the Goursat problem is B\"ar and Wafo's extension (see Theorem 23 in \cite{BarWafo2015}) of a theorem due to H\"ormander \cite{Hormander1990}, which ensures that this can be done from compactly supported data on $\scri^+$. Some care is required, however, since the component $\hat{A}_0^+$ is not compactly supported even if $\hat{A}_2^+ \in C^\infty_c(\scri^+)$. This leads us to solve the Goursat problem near $i^+$ separately, where the solution is pure gauge and we first solve a wave equation for the Maxwell field $\hat{F}_{ab}$ instead. We then recover the potential near $i^+$ using our gauge conditions.

\subsection{Role of timelike conformal symmetry}

As stated in \Cref{thm:main_thm_Minkowski}, in the case of Minkowski space one has, in addition to the standard timelike Killing field $K = \partial_t$, the \emph{conformal} timelike Killing field $K_0 = (t^2 + r^2) \partial_t + 2tr \partial_r$, known as the Morawetz vector field, the generator of `inverted time translations' on $\mathcalboondox{M}$.  Since the Maxwell stress-energy tensor is traceless, $K_0$ also provides a conserved energy which carries different weights, $\mathbf{E}, \, \mathbf{B} \in r^{-1} L^2(\Sigma; \mathbb{R}^3)$. At null infinity, while $K$ becomes tangent to $\scri^+$, $K_0$ is transverse to $\scri^+$, so the energy on $\scri^+$ picks up angular derivatives (see the definition of the space of scattering data $u^{-1} \dot{\mathcal{H}}(\scri^+)$ in \Cref{thm:main_thm_Minkowski}). The overall result is that the spaces of initial and scattering data with respect to $K_0$ are strictly smaller than with respect to $K$, resulting in a stronger scattering theory. Loosely speaking, $\mathscr{S}_{K}$ therefore decomposes into a `direct sum' which contains $\mathscr{S}_{K_0}$ as a factor.

\section*{Acknowledgements}
This work was partly supported by the EPSRC grant [EP/L05811/1]. We thank Gustav Holzegel and Juan Valiente Kroon for useful discussions.


\begin{thebibliography}{99}

\bibitem{TaujanskasNicolas2022} J.-P. Nicolas, G. Taujanskas, Conformal Scattering  of Maxwell Potentials, arXiv:2211.14579, 2022.

\bibitem{ChruscielDelay2003} P. T. Chru\'sciel, E. Delay, On mapping properties of the general relativistic constraints operator in weighted function spaces, with applications, M\'emoires de la Soci\'et\'e Math\'ematique de France {\bf 94}, (2003), 109.

\bibitem{Corvino2000} J. Corvino, Scalar Curvature Deformation and a Gluing Construction for the Einstein Constraint Equations, Commun. Math. Phys. {\bf 214} (1), (2000), 137-189.

\bibitem{CorvinoSchoen2006} J. Corvino, R. M. Schoen, On the Asymptotics for the Vacuum Einstein Constraint Equations, J. Differential Geom. {\bf 73} (2), (2006), 185-217.

\bibitem{ChruscielDelay2002} P. T. Chru\'sciel, E. Delay, Existence of non-trivial, vacuum, asymptotically simple spacetimes, Class. Quantum Gravity {\bf 19} (12), (2002), 3389.

\bibitem{Friedrich86} H. Friedrich, On the existence of $n-$geodesically complete or future complete solutions of Einstein's field equations with smooth asymptotic structure, Commun. Math. Phys. {\bf 107}, (1986), 587-609.

\bibitem{FriedrichSchmidt1987} H. Friedrich, B. G. Schmidt, Conformal geodesics in general relativity, Proc. Roy. Soc. London Ser. A {\bf 414}, (1987), 171-195.

\bibitem{BarWafo2015} C. B\"ar, R. T. Wafo, Initial Value Problems for Wave Equations on Manifolds, Math. Phys. Anal. Geom. \textbf{18} (7), (2015).

\bibitem{Hormander1990} L. H\"ormander, A remark on the characteristic Cauchy problem, J. Funct. Anal. \textbf{93} (2), (1990), 270-277.

\bibitem{Penrose1965} R. Penrose, Zero rest-mass fields including gravitation: asymptotic behaviour, Proc. Roy. Soc. London Ser. A \textbf{284} (1397), (1965), 159-203.

\bibitem{ChristodoulouKlainerman1990} D. Christodoulou, S. Klainerman, The global nonlinear stability of Minkowski space, PMS-41, Princeton University Press, 1994.

\bibitem{HintzVasy2017} P. Hintz, A. Vasy, Stability of Minkowski spaces and polyhomogeneity of the metric, Annals of PDE {\bf 6} (2), (2020).

\bibitem{PenroseRindler2} R. Penrose, W. Rindler, Spinors and space-time Vol. 2: Spinor and twistor methods in space-time geometry, Cambridge University Press, 1986.

\end{thebibliography}
\end{document}